\documentclass{article}
 \title{Stochastic variational principle on riemannian manifolds}
 \author{Gavriel Segre}
 \usepackage{amsmath,amssymb,graphicx}

 \newtheorem{trial definition}{Trial definition}[section]

\begin{document}
 \maketitle
 \newpage
Most of the conceptual appeal of Stochastic Mechanics
\cite{Nelson-85} disappears as soon as one realizes that it may be
ultimatively seen as a corollary of Fukushima's Theorem on
symmetric Dirichlet forms (generalized by Albeverio, Ma and
R\"{o}ckner for non-symmetric Dirichlet forms) applied in the
\textbf{ground-state representation} (sometimes called the
\textbf{Darboux representation})
\cite{Albeverio-Brasche-Rockner-89}, \cite{Ma-Rockner-92},
\cite{Fukushima-Oshima-Takeda-94},\cite{Albeverio-Kondratiev-Rockner-94}.

\medskip

Given an Hilbert space $ {\mathcal{H}} $ and a positive,
symmetric, bilinear form on $ {\mathcal{H}} $ $ {\mathcal{E}} :
D({\mathcal{E}}) \, \times \, D({\mathcal{E}}) \; \rightarrow \;
{\mathbb{R}} $ (a \textbf{symmetric form} from here and beyond)
defined on a dense subset $ D({\mathcal{E}}) \, \times \,
D({\mathcal{E}}) $ of $ {\mathcal{H}} \times {\mathcal{H}} $
(that I will denote as $ (D({\mathcal{E}}) , {\mathcal{E}} ) $)
let us introduce the symmetric form on $ {\mathcal{H}} $:
\begin{align}
  {\mathcal{E}}_{1} & : D({\mathcal{E}}) \, \times \, D({\mathcal{E}}) \;
\rightarrow \; {\mathbb{R}}  \\
   {\mathcal{E}}_{1} ( f , g ) \; & \; := \;  {\mathcal{E}} ( f, g
   ) \,  + \, ( f , g )
\end{align}

\smallskip

Clearly $ (D({\mathcal{E}}) \, , \, {\mathcal{E}}_{1} ) $ is a
\textbf{pre-Hilbert space}.

\smallskip

We will say that the form $  (D({\mathcal{E}}) , {\mathcal{E}} ) $
 is \textbf{closed} if $ (D({\mathcal{E}}) \, , \,
{\mathcal{E}}_{1} ) $ is also an \textbf{Hilbert space}, i.e. if
it is complete in the norm $ \| \cdot \|_{1} \; := \; \sqrt{
{\mathcal{E}}_{1} ( \cdot , \cdot )}$.

\smallskip

We will say that $  (D({\mathcal{E}}) , {\mathcal{E}} ) $  is
\textbf{closable} if it has a closed extension.

In this case we define the \textbf{closure} of $ {\mathcal{E}} $
as the smallest closed extension  $  (D(\bar{{\mathcal{E}}}) ,
\bar{{\mathcal{E}}} ) $ of $  (D({\mathcal{E}}) , {\mathcal{E}} )
$.

\smallskip

The closure condition guarantees the existence of a whole
plethora of bijective correspondences:
\begin{enumerate}
  \item there is a one-to-one correspondence between \textbf{closed symmetric
  forms} and \textbf{positive self-adjoint operators} on $
  {\mathcal{H}} $ according to which  the operator $ ( D(H) , H
  ) $ corresponds the closed symmetric form:
\begin{equation}
  ( D(E) \, := \,
  D(\sqrt{H}) \; , \; E( f , g ) := ( \sqrt{H} \, f \: , \:  \sqrt{H} \,
  g )
\end{equation}
  \item there is a one-to-one correspondence between \textbf{positive self-adjoint
  operators} and \textbf{strongly continuous contraction
  semigroups} on $ {\mathcal{H}} $ according to which the operator $ ( D(H) , H
  ) $ corresponds to the strongly-continuous contraction semigroup
  $ ( T_{t} \, := \, e^{-t H} )_{t \geq 0} $
  \item there is a one-to-one correspondence between \textbf{strongly continuous contraction
  semigroups} and \textbf{strongly continuous contraction
  resolvents} according to which  the strongly continuous
  contraction semigroup $ ( T_{t} )_{ t \geq 0 } $ corresponds to
  the  strongly-continuous contraction-resolvent $ ( G_{\alpha} )_{ \alpha \geq
  0} $:
\begin{equation}
  G_{\alpha} f \; := \; \int_{0}^{\infty} e^{ - \alpha s } \,
  T_{s} f \, ds \; \; \forall f \in {\mathcal{H}}
\end{equation}
\end{enumerate}

\smallskip

The full power of this chain of one-to-one correspondences arises,
anyway, when one consider \textbf{Dirichlet forms}:

in fact, for Dirichlet forms, one can add a link to the  chain of
one-to-one correspondences associating  to a Dirichlet form the
Markovian stochastic process having as
transition-probability-function the integral kernel of the
associated strongly-continuous contraction semigroup.

\smallskip

Supponing that $ {\mathcal{H}} \; = \; L^{2} ( M , \mu ) $ for a
proper Borel measure space $ ( M \, , \, {\mathcal{B}} \, , \,
\mu ) $  will say that the \textbf{closed symmetric form}:  $
(D({\mathcal{E}}) , {\mathcal{E}} ) $ is a \textbf{Dirichlet
form} iff:
\begin{equation}
  {\mathcal{E}} ( f^{\sharp} ,  f^{\sharp} ) \; \leq \;
  {\mathcal{E}}( f , f ) \; \; \forall f \in D ( {\mathcal{E}} )
\end{equation}
where $ f^{\sharp} \; := \; \min ( 1 , \max ( f , 0 )) $.

\smallskip

The \textbf{Dirichlet-condition} of a symmetric form corresponds
to specular conditions on the other links of the chain of
one-to-one correspondences above introduced:

namely one has that:
\begin{equation}
   (D({\mathcal{E}}) , {\mathcal{E}} ) \text{ is Dirichlet } \; \Leftrightarrow \;  ( D(H) , H
  )  \text{ is Dirichlet } \; \Leftrightarrow \; ( T_{t} )_{t \geq
  0} \text{ is markovian } \;
\end{equation}
where a self-adjoint operator $ ( D(A) , A ) $ is said
\textbf{Dirichlet} iff:
\begin{equation}
  ( A f \, , \, \max ( 0 , f - 1 ) ) \; \geq \; 0 \; \; \forall f
  \in D(A)
\end{equation}
while a strongly-continuous contraction semigroup is said
\textbf{markovian} iff:
\begin{equation}
  0 \, \leq f \, \leq \, 1 \; \Rightarrow \; 0 \, \leq T_{t} f \, \leq \,
  1 \; \; \forall f \in {\mathcal{H}}
\end{equation}

\smallskip

Let us now assume that $ ( M , {\mathcal{B} } ) $ is Hausdorff
and that the measure $ \mu $ is $ \sigma-finite$.

\smallskip

We will say that a symmetric form $ {\mathcal{E}} $ is
\textbf{properly associated} with a Markov process $ (X_{t}) _{t
\geq 0} $ on M if $ E_{ \cdot } [ f( X_{t} ) ] $ is an  $
{\mathcal{E}}-$quasi-continuous $ \mu $-version of $ ( e^{-t H} f
) ( \cdot ) $ where $ {\mathcal{E}}$-quasi-continuous means
continuous modulo $ {\mathcal{E}}$- exceptional sets, i.e. modulo
set  $ S \subset M $:
\begin{equation}
  {\mathcal{E}}-cap(S) \; = \; 0
\end{equation}
where the $ {\mathcal{E}} $\textbf{-capacity} of S is defined as:
\begin{equation}
  {\mathcal{E}}-cap(S) \; := \; \inf \{ {\mathcal{E}}_{1} (  f , f ) \, : \, f \in D({\mathcal{E}}) \, , \, f \geq 1 \, , \mu-a.e. \; on \; S  \, \}
\end{equation}
if S is open, and:
\begin{equation}
  {\mathcal{E}}-cap(S) \; := \; \inf \{ {\mathcal{E}}-cap(U) \, : U \; open \; U \supset S  \}
\end{equation}
if S is arbitrary.

\medskip

We will say that a Dirichlet form $  (D({\mathcal{E}}) ,
{\mathcal{E}} ) $ is \textbf{regular} if $ C_{0} (M ) \bigcap
D(\mathcal{E}) $ is dense both in $ (  C_{0} (M) \, , \, \| \cdot
\|_{\infty} ) $ and in $ ( D({\mathcal{E}}) \, , \, \| \cdot
\|_{1}  ) $, where $ C_{0} (M) $ denotes the set of
compact-supported functions on M, $ \| \cdot \|_{\infty} $ denotes
the usual uniform norm, while $ \| \cdot \|_{1} := \sqrt{
{\mathcal{E}}_{1} ( \cdot , \cdot ) } $.

\smallskip

The fundamental Fukushima Theorem grants that if a Dirichlet form
is regular than it is properly associated to a Markov process $ (
X_{t} ) $.

\bigskip

Let us now pass to analyze concretely the more important
application of this general mathematical framework, starting from
the definition  of the brownian motion on a riemannian manifold (M
, g ):

the starting point is the symmetric form:
\begin{equation}
( D ( E ) := C_{0}^{\infty} ( M ) \; , \; E ( f, g ) \; :=  < d f
, d g > )
\end{equation}
where $ C_{0}^{\infty} ( M ) $ denotes the set of
compact-supported smooth functions on M, while $ < \cdot , \cdot
> $ is the \textbf{g - metric induced scalar product} on the space
$  \Gamma ( T^{(0,n)} M  ) $ of the sections of the (0,n)-tensor
bundle (i.e. the space of the n-differential forms on M):
\begin{equation}
  < \omega , \nu > \; := \; \int_{M} \omega \wedge \star \nu \; \;
  \omega , \nu \in \Gamma ( T^{(0,n)} M  )\, , \, n \in {\mathbb{N}}
\end{equation}
where $ \star $ is the Hodge duality operator on (M,g).

\smallskip

The closure $  (D(\bar{{\mathcal{E}}}) , \bar{{\mathcal{E}}} ) $
is a Dirichlet form, usually called the \textbf{heat Dirichlet
form on (M , g )} whose associated positive self-adjoint operator
is precisely the Laplace-Beltrami operator defined on the $ 2^{th}
$ Sobolev space on M:
\begin{equation}
  ( - \Delta_{g} \; := \; d \, d^{\dagger} \, + \,  d^{\dagger} \,
  d \; , \; H^{2}(M) )
\end{equation}
The \textbf{brownian motion on (M,g)} is then defined precisely as
the Markov process on (M,g) properly associated to the
\textbf{heat Dirichlet form on (M , g )}.

\smallskip

Let us now  modify a bit the scenario introducing the Hilbert
space:
\begin{equation}
  {\mathcal{H}} \; := \;  L^{2} ( M , d \mu_{g} )
\end{equation}
and  associating to a generic \textbf{wave-function} $ \psi \in
{\mathcal{H}}  $ the symmetric form:
\begin{equation}
( D ( E_{\psi} ) := C_{0}^{\infty} ( M ) \; , \; E_{\psi} ( f, g )
\; := \int_{M} | \psi | ^{2} d f \, \wedge \, \star d g )
\end{equation}
on the Hilbert space:
\begin{equation}
  {\mathcal{H}}_{\psi} \; := \; L^{2} ( M , | \psi | ^{2} \, d
  \mu_{g} )
\end{equation}
Since its closure $ (D(\bar{{\mathcal{E_{\psi}}}}) ,
\bar{{\mathcal{E_{\psi}}}} ) $ is a Dirichlet form, the
Fukushima's Theorem allows to associate to the
\textbf{wave-function} $ \psi $ the \textbf{markovian stochastic
process} $ x_{\psi} $  over (M , g ) properly associated to $
(D(\bar{{\mathcal{E_{\psi}}}}) , \bar{{\mathcal{E_{\psi}}}} ) $.

For example, in the particular case in which $ ( M , g ) $ is the
one-dimensional euclidean space $ ( {\mathbb{R}} \, , \, \delta )
$ while:
\begin{equation} \label{eq:gaussian wave-function on a one dimensional euclidean manifold}
  \psi (y) \; := \; \frac{1}{ \pi ^{\frac{1}{4}}} \, e ^{ - \frac{ y^{2} }{2}}
\end{equation}
is the ground-state wave-function of the harmonic-oscillator's
hamiltonian,  $ x_{\psi} $ is the Ornstein-Uhlenbeck stochastic
process over $ ( {\mathbb{R}} \, , \, \delta ) $.

Returning to the general case let us denote by $ \hat{H}_{\psi} $
the positive self-adjoint operator over $ {\mathcal{H}}_{\psi} $
associated to the Dirichlet form $ (D(\bar{{\mathcal{E_{\psi}}}})
, \bar{{\mathcal{E_{\psi}}}} ) $.

One has that $ \psi $ is the ground-state eigenfunction of $
\hat{H}_{\psi} $, correponding to the zero eigenvalue.

It easily follows that $ \psi $ is the solution of the
variational problem:
\begin{align}
  \delta S_{1} ( \psi )  \; &  = \; 0 \\
  \delta^{2} S_{1} ( \psi )  \; &  > \; 0
\end{align}
where  $ S_{1} \: : \: {\mathcal{H}} \, \mapsto \, {\mathbb{R}} $
is the functional:
\begin{equation}
  S_{1} ( \psi ) \; := \; \frac{ < \psi | \hat{H}_{\psi} | \psi >  }{ < \psi | \psi > }
\end{equation}
Denoted by $ {\mathcal{M}} ( M , g ) $ the set of all the
markovian stochastic processes over ( M , g ) let us introduce
the functional $ S_{2} \: : \: {\mathcal{H}} \, \mapsto \,
{\mathbb{R}} $ defined as:
\begin{equation}
  S_{2} ( x_{\psi} ) \; := \; S_{1} ( \psi )
\end{equation}
One has then that, obviously,  $ x_{\psi} $ is the solution of
the stochastic action principle:
\begin{align}
  \delta S_{2} ( x_{\psi} )  \; &  = \; 0 \\
  \delta^{2} S_{2} ( x_{\psi} )  \; &  > \; 0
\end{align}
Let us now observe that such a stochastic variational principle
is different and incompatible with Guerra's way of formulating the
stochastic action principle on manifolds \cite{Guerra-88a},
\cite{Guerra-Aldrovandi-Dohrn-89},\cite{Guerra-Aldrovandi-Dohrn-90}
as we will know show:

let us introduce the following one-parameter family of wave
functions belonging to the Hilbert space $ {\mathcal{H}} \; :=
\;  L^{2} ( M , d \mu_{g} ) $:
\begin{equation} \label{eq:gaussian wave-function on an arbitrary riemannian manifold}
  \psi_{p}(q) \; := \; N(p) e^{ \frac{ (d_{g}(p,q))^{2}}{2}} \; \;
  p \in M
\end{equation}
where $ d_{g}(p,q) $ is the geodesic distance among the points p
and q of (M,g), while N(p) is a normalization factor such that:
\begin{equation}
  \| \psi_{p} \|_{{\mathcal{H}}}^{2} \; = \; 1 \; \; \forall p \in M
\end{equation}
Let us now observe that in the particular case in which $ (M , g)
$ is the one-dimensional euclidean space $ ( {\mathbb{R}} \, , \,
\delta ) $ and $ p := 0 $ the wave-function  of
eq.\ref{eq:gaussian wave-function on an arbitrary riemannian
manifold} reduces to the one-dimensional harmonic-oscillator's
ground-state's wave-function of eq.\ref{eq:gaussian wave-function
on a one dimensional euclidean manifold} and conseguentially, the
markovian process $ x_{\psi} $  reduces to the one-dimensional
Ornstein-Uhlenbeck process.

This doesn't occurs, instead in the Guerra's approach  to the same
situation whose starting point is the ill-defined construction of
a stochastic action functional that is:
\begin{itemize}
  \item expressed in terms of local coordinates but not invariant
under change of local coordinates
  \item dependent among the totally arbitrary choice of an affine
  connection that is not fixed to be the Levi-Civita connection of
  (M ,g) and whose variation is, in a totally inconsistent
  way, claimed to correspond to a sort of gauge transformation
  \item depends critically upon an ad-hoc regularization
  procedure, resulting in the more possibly arbitary renormalized
  stochastic action
\end{itemize}
The absolute mathematical inconsistence of such an approach is
definitively proved by the own fact that, in the one-dimensional
euclidean limit, the solution of the stochastic variational
equation it gives rise to is not the Ornstein-Uhlenbeck process.

\end{document}